\documentclass[10pt,twocolumn]{article}

\usepackage[utf8]{inputenc}
\usepackage{amsmath}
\usepackage{amssymb}
\usepackage{titlesec}
\usepackage{cite}
\usepackage{booktabs}
\usepackage{graphicx}
\usepackage[colorlinks=false]{hyperref}
\usepackage[format=plain,labelfont=it]{caption}
\usepackage[left=1.5cm,right=1.5cm,top=2cm,bottom=2cm]{geometry}
\usepackage{xcolor}
\usepackage[tight, nice]{units}

\titleformat{\section}{\centering\normalfont\scshape}{\Roman{section}.}{5pt}{}
\titleformat{\subsection}{\normalfont\it}{\Alph{subsection}.}{5pt}{}
\titleformat{\subsubsection}{\normalfont\it}{\hspace{4mm}\arabic{subsubsection})}{5pt}{}

\newcommand\infoFootnote[1]{%
	\begingroup
	\renewcommand\thefootnote{}\footnotetext{#1}%
	\addtocounter{footnote}{-1}%
	\endgroup}

\newcommand{\preprintNoteIFACAccepted}[2]{
	$\copyright$ {#1}.
	This paper is a \textbf{preprint} of a contribution to the {#2}.
	This work is published (\url{https://doi.org/10.1016/j.ifacol.2024.09.004}) under a Creative Commons Licence CC-BY-NC-ND.
}

\hypersetup{
	colorlinks  = true, 
	linktocpage = true, 
	allcolors = {black}, 
	breaklinks  = true, 
}

\newcommand{\R}{\mathbb{R}}

\newcommand{\Ab}{\boldsymbol{A}}

\newcommand{\Cb}{\boldsymbol{C}}
\newcommand{\Db}{\boldsymbol{D}}

\newcommand{\Ib}{\boldsymbol{I}}

\newcommand{\Wb}{\boldsymbol{W}}

\newcommand{\bb}{\boldsymbol{b}}

\newcommand{\fb}{\boldsymbol{f}}

\newcommand{\ub}{\boldsymbol{u}}
\newcommand{\vb}{\boldsymbol{v}}
\newcommand{\wb}{\boldsymbol{w}}
\newcommand{\xb}{\boldsymbol{x}}
\newcommand{\yb}{\boldsymbol{y}}

\newcommand{\zerob}{\boldsymbol{0}}

\newcommand{\Sigmab}{\boldsymbol{\Sigma}}
\newcommand{\xib}{\boldsymbol{\xi}}

\newcommand{\zetab}{\boldsymbol{\zeta}}
\newcommand{\xbh}{\hat{\boldsymbol{x}}}
\newcommand{\ybh}{\hat{\boldsymbol{y}}}

\newcommand{\Nc}{\mathcal{N}}

\title{\vspace{-2mm}\bf MPC using mixed-integer programming \\ for aquifer thermal energy storages}
\author{Johannes van Randenborgh and Moritz Schulze Darup\vspace{2mm}}
\date{}

\begin{document}
	
	\maketitle
	
	\textbf{\textit{Abstract}.} {\bf Aquifer thermal energy storages (ATES) are used to temporally store thermal energy in groundwater saturated aquifers.
		Typically, two storages are combined, one for heat and one for cold, to support heating and cooling of buildings.
		This way, the use of classical fossil fuel-based heating, ventilation, and air conditioning can be significantly reduced.
		Exploiting the benefits of ATES beyond ``seasonal'' heating in winter and cooling in summer as well as meeting legislative restrictions requires sophisticated control.
		We propose a tailored model predictive control (MPC) scheme for the sustainable operation of ATES systems, which mainly builds on a novel model and objective function.
		The new approach leads to a mixed-integer quadratic program.
		Its performance is evaluated on real data from an ATES system in Belgium.}
	\infoFootnote{J. van Randenborgh and M. Schulze Darup are with the \href{https://rcs.mb.tu-dortmund.de/}{Control and~Cyber-physical Systems Group}, Faculty of Mechanical Engineering, TU Dortmund University, Germany. E-mails:  \{first-name\}.\{last-name\}@tu-dortmund.de. \vspace{0.5mm}}
	\infoFootnote{\hspace{-1.5mm}$^\ast$
		\preprintNoteIFACAccepted{van Randenborgh and Schulze Darup}{8th IFAC Conference on Nonlinear Model Predictive Control}
		
	}
	
	\section{Introduction}\label{sec:Introduction}
	
	Members of the \cite{UNFCCC.2022} decided to strengthen efforts to limit global temperature increase to $\unit[1.5]{^\circ C}$.
	According to the \cite{InternationalEnergyAgency.2023}, building operation has been responsible for $26\%$ of global energy-related emissions in 2022.
	In particular, the report states that heating, ventilation, and air conditioning (HVAC) of buildings produces in total $4.1 \mathrm{Gt}$ of direct and indirect CO$_\textbf{2}$ emissions, offering tremendous potential for reducing greenhouse gas emission.
	Following the political interests, industry and research already present a wide range of HVAC technologies to replace fossil fuel-based heating and cooling for the building sector.
	Among these, underground thermal energy storages (UTES) stand out for their large energy storage capacities with low operational cost \cite{Lee.2013}.
	Depending on the storage type, UTES can be further subdivided into borehole, cavern, or aquifer thermal energy storages (ATES).
	All of those come with specific advantages and requirements, e.g., for geological properties.
	Here, we focus on ATES since they provide large heat transfer and storage capacities \cite{Lee.2013}.
	Both features are suitable for densely populated urban regions.
	
	\subsection{Principles of ATES operation}\label{sec:Principles_of_ATES_operation}
	
	The basic idea of ATES systems is to inject heat from summer into an aquifer and to use it during winter for heating.
	A second aquifer stores cold in winter for cooling in summer.
	However, in application, injection of heat and cold alternates frequently, e.g., during spring and autumn forcing ATES systems to dynamically interfere with buildings.
	Storing energy into aquifers results in local temperature changes of groundwater and rock perturbing present ambient temperature.
	ATES systems comprise at least one aquifer for cold and for heat with possible additional sub-components such as, e.g., heat exchangers (HX) or heat pumps.
	Energy injection and production from the subsurface are performed simultaneously.
	In summer, for instance, cold is produced by the second aquifer and exchanged with the building.
	There, the cold heats up by the exchange process, becomes heat, and is injected into the first aquifer.
	The transported heat and cold are carried by groundwater, which is injected and extracted in equal amounts into the aquifers.
	ATES systems have three operational modes: heating, storing (or inactivity), and cooling, which depend on the pumping flow direction between the aquifers.
	They contribute to the total energy demands of buildings and lower their CO$_{2}$ emissions by reducing operations of additionally installed fossil fuel-based HVAC technology.
	Further, the operational costs are lower compared to gas heating \cite{Desmedt.2007}.
	
	Optimal operation of ATES systems is crucial for achieving significant and dynamic contributions to the energy demand of buildings.
	However, the subsurface's vitality must be maintained with a focus on preserving potable groundwater sources.
	For instance, large temperature changes in the subsurface may trigger bio-chemical reactions or promote clogging \cite{Bonte.2015}.
	Despite the possible consequences of temperature change in the ground, profound experimental field studies on the effect of temperature change on chemical or biological reactions are not yet performed by research \cite{Hartog.2013}.
	As a consequence, the operation of ATES underlies prohibitive legislative restrictions in Germany and the Netherlands.
	For instance, the VDI~4630 regulates the stored energy in ATES such that equal amounts of heat and cold are stored in the subsurface within a certain time horizon \cite[Part 3]{VDI4640.2020}.
	Similar rules also apply to other UTES systems, such as borehole heat exchangers that store heat and cold in the ground.
	Further, persistent energy unbalances of ATES system operation lead to low viability and depletion of the aquifer storages, which may ultimately result in the cessation of operation.
	A prominent example is the Reichstag building of the German parliament, where the operation of the ATES system was ceased due to ongoing unbalanced operation \cite{Fleuchaus.2020}.
	
	In this context, assuming a balanced heat and cold demand of buildings is not sufficient for guaranteeing energy balanced operation of ATES systems.
	Studies have shown that the energy demand of buildings mainly depends on their construction, use, and weather, regardless of the length of winter and summer period \cite{Beernink.2022}.
	As a result, provided heat and cold of ATES systems are not balanced for purely heat demand-driven operations, which results in the necessity of controlling the power output of ATES systems accordingly \cite{Beernink.2022, Hoving.2017}.
	
	\subsection{Control approaches for ATES}\label{sec:Control-approaches-for-ATES}
	
	Conservative control of ATES systems, known for seasonal operations, builds on elementary or even manual control \cite{Vanhoudt.2011}.
	However, once a more flexible and efficient operation of ATES systems is desired, sophisticated control schemes are required.
	As a result, literature already presents a variety of MPC schemes for ATES, which we briefly summarize next.
	
	Basic battery-like first principle linear models of ATES' storing statuses are presented by \cite{Rostampour.2016c}.
	Heat loss in the ground is considered by a time-invariant constant.
	Moreover, the ground's temperature is assumed to be spatially and temporally constant.
	A further publication by \cite{Rostampour.2017}, based on a similar model, presents an MPC approach for smart thermal grids, capable of constantly monitoring ATES' power delivery.
	Perfect mixing of injected enthalpy is assumed.
	Addressing the need for sustainable operation of ATES, the authors enforce balanced operation with linear inequality constraints at the end of the prediction horizon.
	Uncertainties of the building's heat demand are considered by a robust-randomized approach.
	
	A case study by \cite{Hoving.2017} demonstrates the capabilities of MPC to achieve energy balance of ATES systems.
	The authors present an empirical model for ATES systems tuned on data from a plant in Utrecht, the Netherlands.
	It is described that the building demand is unbalanced with a larger cooling than heating demand.
	To act against unbalanced operations of the ATES system, a sub-component, an air handling unit, can additionally add cold to the cold aquifer for regeneration.
	The MPC scheme tracks a predefined energy balance trajectory and is only employed during winter season controlling the regeneration with the air handling unit.
	
	Being capable of considering transient ground temperatures of ATES, a control oriented modeling framework with a nonlinear first principle mixed-logical dynamical model is proposed by \cite{Rostampour.2016}.
	Perfect mixing of enthalpy is still assumed and lumped loss coefficients approximate heat loss based on simulations.
	The authors indicate a large computational cost solving derived nonlinear optimization problem that is mixed-integer multi-dimensional polynomials nonlinear programming.
	A follow-up publication by \cite{Rostampour.2017b} lowers the computational effort by pre-defining operational modes depending on outside temperatures.
	
	\subsection{Challenges and contributions}\label{sec:challenges-and-contributions}
	
	The given control approaches in Section \ref{sec:Control-approaches-for-ATES} make use of assumptions that do not allow for consideration of temporal and spatial ground temperature changes.
	However, such changes are present, due to, for instance, ambient groundwater fluxes, anisotropic material properties, weather, and ATES pumping actions \cite{Bloemendal.2018, Bonte.2015}.
	Hence, the MPC schemes might have trouble predicting extraction temperatures of the aquifers and, consequently, the delivered power of the ATES system to the building.
	This publication follows up on that with a more detailed model for capturing the energy transport and distribution in the aquifers.
	We believe that this can contribute to an improved operation of ATES and other HVAC technology.
	
	Furthermore, the predefinition of operational modes \cite{Rostampour.2017b} may force ATES systems to operate against sustainable operation principles; that is, the injection of warm fluid into the cold aquifer.
	The tailored MPC scheme, described in Section~\ref{sec:A-tailord-MPC-scheme}, is able to decide, based on the objective function, which operational mode should be used.
	The objective function combines the aim for sustainable and cost-efficient operation of ATES systems.
	Operational priorities may be tuned by the user.
	
	\subsection{Organization}\label{sec:Organization}
	
	Section \ref{sec:A-novel-model-for-ates-systems} presents the model derivation of the ATES system.
	Next, proposed MPC scheme is presented in Section \ref{sec:A-tailord-MPC-scheme}.
	To test the findings of the previous sections, a cursory numerical study is discussed in Section \ref{sec:Numerical-Study}.
	Finally, conclusions and the outlook are given in Section \ref{sec:Conclusions-and-outlook}.
	
	\section{A novel model for ATES systems}\label{sec:A-novel-model-for-ates-systems}
	
\begin{figure}[t]
		\centering
		\includegraphics[trim={0, 130, 480, 0}, clip, width=0.4\textwidth]{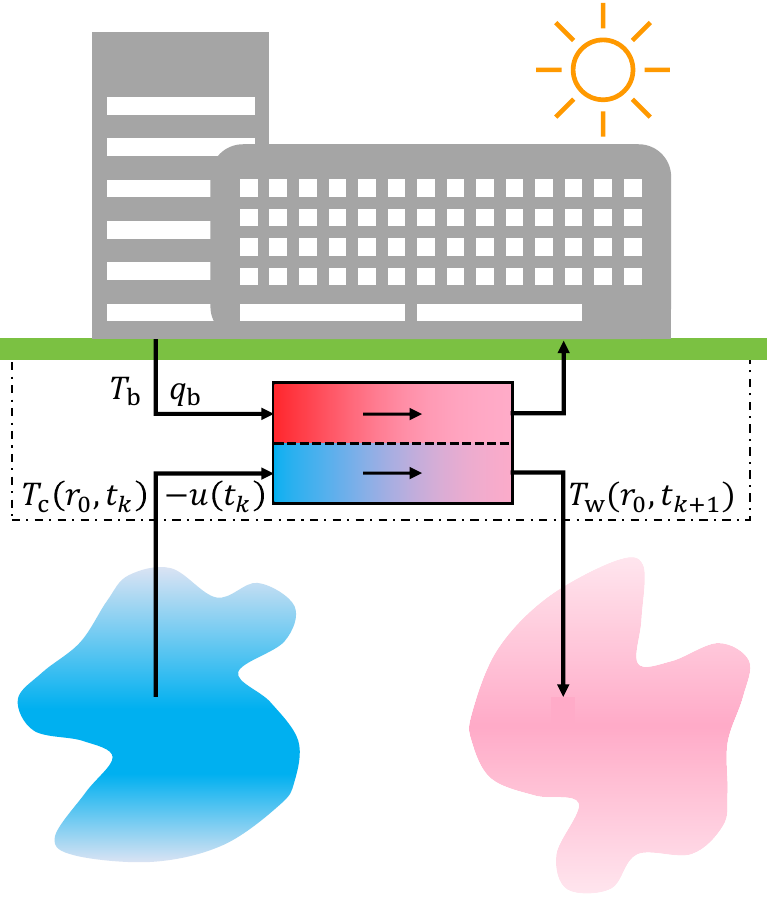}
		\caption{Schematic illustration of the ATES system in cooling mode (i.e., $u(t_k)<0$).}
		\label{fig:ATES_principle}
	\end{figure}

	Our model builds on discretized partial differential equations (PDEs) describing the temperature profiles in the two aquifers and a simplified model of a HX connecting the aquifers and building (see Fig.~\ref{fig:ATES_principle}).
	The central control variable~$u(t)$ is the flow rate from one aquifer to the other, where $u(t)>0$ refers to the heating mode (i.e., flow from the warm to the cold aquifer).
	Consequently, $u(t)=0$ reflects storing and $u(t)<0$ refers to cooling.
	Next, we describe the models of the sub-systems and combine them to a piecewise affine (PWA) system model in Section~\ref{sec:mixed-dynamical-ATES-model}.
	
	\subsection{Temperature profiles in ATES}\label{sec:ATES}
	
\begin{figure}[t]
		\centering
		\includegraphics[trim={0 300 700 0}, clip, width=0.4\textwidth]{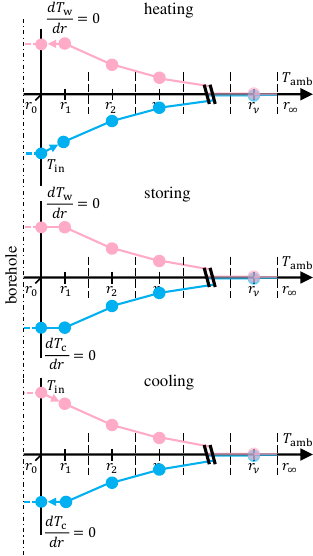}
		\caption{Illustration of boundary conditions for and discretization of the PDE~\eqref{eq:T_PDE_simplified} for the warm and cold aquifer depending on the different operation modes of the ATES system. Arrows indicate flow directions.}
		\label{fig:Boundary_Conditions}
	\end{figure}

	Assuming radial symmetry, temperature profiles in groundwater saturated aquifers can be described by the 1-D parabolic PDE 
	\begin{equation}
		\label{eq:T_PDE_nonlinear}
		c_{\mathrm{a}} \frac{{\partial T(r,t)}}{{\partial t}} =  \frac{\lambda}{r} \frac{\partial}{\partial r} \left( r \frac{\partial T(r,t)}{\partial r}\right) - c_{\mathrm{w}} v(r,t) \frac{\partial T(r,t)}{\partial r},
	\end{equation}
	where $c_{\mathrm{a}}$ and  $c_{\mathrm{w}}$ denote the specific volumetric heat capacity of the aquifer and water, respectively, $\lambda$ is the heat conduction coefficient of the subsurface, and $v(r,t)$ is the radial flow velocity \cite{Anderson.2005}.
	The PDE's solution is the fluid temperature~$T(r,t)$ in the aquifer as a function of the radial distance~$r$ to the borehole and time~$t$.
	The PDE \eqref{eq:T_PDE_nonlinear} is based on the equation of energy for pure Newtonian fluids.
	More precisely, the rate of change of internal energy (on the left-hand side) is determined by the interplay between conduction and convection (first and second term on the right-hand side).
	While both physical effects act simultaneously, temperature change by advection is more dominant \cite{Kim.2010}.
	As a consequence, we slightly simplify the term for convection in \eqref{eq:T_PDE_nonlinear} without compromising significant accuracy.
	First, we assume that the flow velocity~$v(r,t)$ is determined by the volume flow~$q(t)$ into (or, for $q(t)<0$, out of) the aquifer via
	\begin{equation*}
		v(r,t) = \frac{1}{2\pi r l} q(t),
	\end{equation*}
	where $l$ reflects the length of the borehole's filter segment, which allows fluid exchange with the aquifer.
	Notably, by construction, we have $q(t)=u(t)$ for the cold aquifer and $q(t)=-u(t)$ for the warm one.
	Second, we fix the temperature gradient for the convection term during predictions (within the MPC scheme).
	Hence, \eqref{eq:T_PDE_nonlinear} simplifies to 
	\begin{equation}
		\label{eq:T_PDE_simplified}
		c_{\mathrm{a}} \frac{{\partial T(r,t)}}{{\partial t}} =\frac{\lambda}{r} \frac{\partial}{\partial r} \left( r \frac{\partial T(r,t)}{\partial r}\right) -   \frac{c_{\mathrm{w}}}{2\pi r l}  \frac{\partial T(r,t^\circ)}{\partial r} q(t),
	\end{equation}
	where $t^\circ$ reflects the initial time for a prediction.
	The temperature profiles in the aquifers are further characterized by boundary conditions.
	For both aquifers, we assume that the temperature far from the borehole equals some ambient temperature~$T_{\mathrm{amb}}(t)$.
	In our model, this is reflected by the Dirichlet condition $T(r_{\infty},t)=T_{\mathrm{amb}}(t^\circ)$ for some $r_\infty \gg r_{0}$, where $T_{\mathrm{amb}}(t^\circ)$ indicates that the ambient temperature is fixed during predictions.
	Boundary conditions at the borehole (i.e., at~$r_{0}$) differ for the warm and cold aquifer and for the operation mode of the ATES system.
	More precisely, whenever fluid is injected into an aquifer (which happens at the cold aquifer during heating and at the warm one during cooling), we consider the Dirichlet condition $T(r_{0},t)=T_{\mathrm{inj}}(t)$.
	In contrast, when fluid is extracted from or stored in an aquifer, the Neumann condition
	\begin{equation*}\label{eq:Boundary_Condition_Neumann}
		\frac{\partial T(r_{0},t)}{\partial t} = 0
	\end{equation*}
	applies.
	The acting boundary conditions for the different modes are illustrated (on the left) in Figure~\ref{fig:Boundary_Conditions}.
	Each set of boundary conditions ensures a unique solution of the PDE as shown by \cite[Thm.~8]{Protter.1984}.
	
	Now, to build our ATES prediction model, we consider one PDE~\eqref{eq:T_PDE_simplified} for the warm and one for the cold aquifer, where we denote the corresponding temperature profiles with $T_{\mathrm{w}}$ and $T_{\mathrm{c}}$, respectively.
	We then apply standard spatial and temporal discretizations of the PDEs (see, e.g., \cite[Chap.~9.1]{Schafer.2022}.
	Here, the spatial domain of each aquifer is divided into $\nu$ cells (as indicated in Fig.~\ref{fig:Boundary_Conditions}).
	The temporal discretization builds on a sampling time $\Delta t$.
	Without giving further details on the finite difference scheme (for brevity), we note that the resulting temperature dynamics can be formulated as
	\begin{equation}\label{eq:System_Dynamics_Extraction}
		\xb_{\mathrm{c}}(k\!+\!1) = \Ab_{\mathrm{c}}^{\mathrm{ex}}(t^\circ) \xb_{\mathrm{c}}(k) + \bb_{\mathrm{c}}^{\mathrm{ex}}(t^\circ) u(k) + \fb_{\mathrm{c}}^{\mathrm{ex}}(t^\circ)
	\end{equation}
	for the cold aquifer subject to fluid extraction, where $k$ reflects the point of time $t_{k}:=t^\circ+k\Delta t$ and where 
	\begin{equation*}
		\xb_{\mathrm{c}}(k):=
		\begin{pmatrix}
			T_{\mathrm{c}}(r_{0},t_{k}) & T_{\mathrm{c}}(r_1,t_{k}) & \dots & T_{\mathrm{c}}(r_\nu,t_{k})
		\end{pmatrix}^\top \in \R^{\nu+1}
	\end{equation*}
	collects the temperatures at $r_{0}$ and at the midpoints~$r_i$ of the various cells (see Fig.~\ref{fig:Boundary_Conditions}).
	Note that the parameters $\Ab_{\mathrm{c}}^{\mathrm{ex}}(t^\circ)$, $\bb_{\mathrm{c}}^{\mathrm{ex}}(t^\circ)$, and $\fb_{\mathrm{c}}^{\mathrm{ex}}(t^\circ)$ are updated for every prediction as indicated by the argument $t^\circ$, which we omit for compactness in the following.
	Now, dynamics analog to~\eqref{eq:System_Dynamics_Extraction} result for the warm aquifer subject to fluid extraction.
	We denote the corresponding parameters with $\Ab_{\mathrm{w}}^{\mathrm{ex}}$, $\bb_{\mathrm{w}}^{\mathrm{ex}}$, and $\fb_{\mathrm{w}}^{\mathrm{ex}}$.
	The cases of fluid injection are more delicate as at that moment the temperature at the borehole depends on the output temperature of the HX.
	This dependency is clarified next.
	
	\subsection{Heat exchanger}\label{sec:Heat_exchanger}
	
	As indicated in Figure~\ref{fig:ATES_principle}, the warm and cold aquifer are connected via an HX.
	For the illustrated flow direction and under the assumption that no thermal losses occur in the boreholes and pipes, the water on the ATES side enters the HX with temperature~$T_{\mathrm{c}}(r_{0},t)$.
	Then, heat is collected from the building side and the warmed water is injected into the warm aquifer.
	Due to the acting Dirichlet condition, the temperature at the HX's outlet determines $T_{\mathrm{w}}$ at $r_{0}$.
	It remains to specify this temperature in relation to the inlet temperature and the conditions on the building side.
	Under the assumption of a constant inlet temperature~$T_{\mathrm{b}}$ and volume flow~$q_{\mathrm{b}}$ on the building side, we find the nonlinear relation 
	\begin{equation} \label{eq:nonLinRelationT_HX}
		T_{\mathrm{w}}(r_{0},t_{k+1}) = \frac{q_{\mathrm{b} } }{ q_{\mathrm{b}}  - u(t_{k}) }  \left( T_{\mathrm{b}} - T_{\mathrm{c}}(r_{0},t_{k})\right)  + T_{\mathrm{c}}(r_{0},t_{k}) 
	\end{equation}
	for an idealized cocurrent HX (with infinitely large heat transfer coefficients and surface areas) according to \cite[Chap.~C1]{Roetzel.2010}.
	Using a first order Taylor expansion of~\eqref{eq:nonLinRelationT_HX} for predictions eventually leads to an affine relation of the form
	\begin{equation} \label{eq:affineRelationT_HX}
		T_{\mathrm{w}}(r_{0},t_{k+1}) = a_{\mathrm{c}}(t^\circ) T_{\mathrm{c}}(r_{0},t_{k}) + b_{\mathrm{c}}(t^\circ) u(t_{k}) + f_{\mathrm{c}}(t^\circ), 
	\end{equation}
	where the index ``${\mathrm{c}}$'' refers to cooling mode.
	An analog relation with (scalar) parameters $a_{\mathrm{h}}$, $b_{\mathrm{h}}$, and $f_{\mathrm{h}}$ results for the heating mode, where water with temperature~$T_{\mathrm{w}}(r_{0},t_{k})$ is extracted from the warm aquifer, cooled down in the HX, and injected into the cold aquifer with the temperature~$T_{\mathrm{c}}(r_{0},t_{k+1})$.
	
	\subsection{Combined model for ATES system}\label{sec:mixed-dynamical-ATES-model}
	
	It remains to combine the models for the aquifers and the HX.
	The three modes of the ATES system and the affine structures of the modules~\eqref{eq:System_Dynamics_Extraction} and \eqref{eq:affineRelationT_HX} result in
	a piecewise affine (PWA) model of the form 
	\begin{equation}\label{eq:PWA-Model}
		\xb(k\!+\!1) = 
		\begin{cases} 
			\Ab_{1} \, \xb(k) + \bb_{1} \, u(k) + \fb_{1} & \text{if } u(k) >0, \\
			\Ab_{2} \, \xb(k) + \fb_{2} & \text{if } u(k) = 0, \\
			\Ab_{3} \, \xb(k) + \bb_{3} \, u(k) + \fb_{3} & \text{if } u(k) < 0,
		\end{cases}
	\end{equation}
	where the sign of $u(k)$ determines the case and where state~$\xb \in \R^{n}$ reflects the concatenation of the discretized temperature profiles~$\xb_{\mathrm{w}}$ and $\xb_{\mathrm{c}}$, i.e., 
	\begin{equation*}
		\xb(k)^\top:=
		\begin{pmatrix}
			\xb_{\mathrm{w}}(k)^\top & \xb_{\mathrm{c}}(k)^\top 
		\end{pmatrix}.
	\end{equation*}
	The number of states~$n$ is defined by $n = 2 (\nu + 1)$.
	Based on these specifications, we derive
	\begin{align*} 
		\Ab_{1} & = 
		\begin{pmatrix}
			\Ab_{\mathrm{w}}^{\mathrm{ex}} & \zerob \\
			\begin{bmatrix}
				a_{\mathrm{h}} & \zerob
			\end{bmatrix}  & \zerob \\
			\zerob & \Ab_{\mathrm{c}}^{\mathrm{inj}}
		\end{pmatrix}, & 
		\bb_{1} & = 
		\begin{pmatrix}
			\bb_{\mathrm{w}}^{\mathrm{ex}} \\
			b_{\mathrm{h}}\\
			\bb_{\mathrm{c}}^{\mathrm{inj}}
		\end{pmatrix}, & 
		\fb_{1} & = 
		\begin{pmatrix}
			\fb_{\mathrm{w}}^{\mathrm{ex}} \\
			f_{\mathrm{h}}\\
			\fb_{\mathrm{c}}^{\mathrm{inj}}
		\end{pmatrix}, \\
		\Ab_{2} & = 
		\begin{pmatrix}
			\Ab_{\mathrm{w}}^{\mathrm{ex}} & \zerob\\
			\zerob & \Ab_{\mathrm{c}}^{\mathrm{ex}}
		\end{pmatrix}, & 
		&  & 
		\fb_{2} & = 
		\begin{pmatrix}
			\fb_{\mathrm{w}}^{\mathrm{ex}} \\
			\fb_{\mathrm{c}}^{\mathrm{ex}}
		\end{pmatrix}, \\
		\Ab_{3} & = 
		\begin{pmatrix}
			\zerob &  \begin{bmatrix}
				a_{\mathrm{c}} & \zerob
			\end{bmatrix} \\
			\Ab_{\mathrm{w}}^{\mathrm{inj}} & \zerob \\
			\zerob& \Ab_{\mathrm{c}}^{\mathrm{ex}}
		\end{pmatrix}, & 
		\bb_{3} & = 
		\begin{pmatrix}
			b_{\mathrm{c}}\\
			\bb_{\mathrm{w}}^{\mathrm{inj}} \\
			\bb_{\mathrm{c}}^{\mathrm{ex}}
		\end{pmatrix}, & \text{and }
		\fb_{3} & = 
		\begin{pmatrix}
			f_{\mathrm{c}}\\
			\fb_{\mathrm{w}}^{\mathrm{inj}} \\
			\fb_{\mathrm{c}}^{\mathrm{ex}}
		\end{pmatrix},
	\end{align*}
	where $\Ab_{\mathrm{c}}^{\mathrm{inj}} \in \R^{\nu \times (\nu+1)}$ and $\bb_{\mathrm{c}}^{\mathrm{inj}},\fb_{\mathrm{c}}^{\mathrm{inj}}\in \R^\nu$ specify the relation of $\xb_{\mathrm{c}}(k)$ and $\begin{pmatrix}T_{\mathrm{c}}(r_1,t_{k+1}) & \dots & T_{\mathrm{c}}(r_\nu,t_{k+1})\end{pmatrix}^\top \in \R^{\nu}$ during fluid injection into the cold aquifer (similar to~\eqref{eq:System_Dynamics_Extraction}).
	Analogously, $\Ab_{\mathrm{w}}^{\mathrm{inj}} \in \R^{\nu \times (\nu+1)}$ and $\bb_{\mathrm{w}}^{\mathrm{inj}},\fb_{\mathrm{w}}^{\mathrm{inj}}\in \R^\nu$ handle injection for the warm aquifer.
	Recall that the missing $T_{\mathrm{c}}(r_0,t_{k+1})$ or $T_{\mathrm{w}}(r_0,t_{k+1})$ is specified by the HX in both cases.
	
	\section{A tailored MPC scheme}\label{sec:A-tailord-MPC-scheme}
	
	Next, we present an MPC scheme tailored for the derived model and the specified control goals.
	As a starting point, we consider the classical optimal control problem (OCP)
	\begin{equation} \label{eq:OCP}
		V_{N}(\xb^\circ)  :=  \min_{\xb_{N}, \ub_{N}} J(\xb_{N}, \ub_{N}) 
	\end{equation}
	subject to the system dynamics~\eqref{eq:PWA-Model}, the initial condition $\xb(0)=\xb^{\circ}$, and the box constraints
	\begin{equation*}
		\underline{\ub}_{N} \leq \ub_{N} \leq \overline{\ub}_{N} \quad \text{and} \quad \underline{\xb}_{N} \leq \xb_{N} \leq \overline{\xb}_{N}.
	\end{equation*}
	Here, $\ub_{N}$ and $\xb_{N}$ stand for the predicted sequences
	\begin{align*}
		\ub_{N} := 
		\begin{pmatrix}
			u(0) \, \dots \, u(N-1)
		\end{pmatrix}^{\top} \!\!\!\! \text{ and }
		\xb_{N}^\top := 
		\begin{pmatrix}
			\xb^{\top}(0) \, \dots \, \xb^{\top}(N)
		\end{pmatrix}.
	\end{align*}
	While the form of the OCP is classical, some elements need further attention.
	First, due to the PWA model~\eqref{eq:PWA-Model}, the OCP~\eqref{eq:OCP} (including the constraints) results in a mixed-integer (MI) problem (see, e.g., \cite{Bemporad.1999}).
	Moreover, we need to specify the cost function~$J$ according to our control goals.
	Finally, implementing the MPC scheme requires (an estimation of) the current system state~$\xb^{\circ}$.
	Since measurements are only available for the temperatures in the boreholes, we design a suitable state observer in Section \ref{sec:nonlinear-state-observer}. 
	
	\subsection{Objective function}\label{sec:Objective-function}
	
	We consider the objective function
	\begin{multline}
		\label{eq:objective_function}
		J(\xb_{N}, \ub_{N}) = \\ 
		\ell_{\mathrm{e}}(\xb(0), \xb(N)) + \!\sum_{k=0}^{N-1} \ell_{\mathrm{d}}(\xb(k+1), \xb(k)) + \ell_{\mathrm{u}}(u(k))
	\end{multline} 
	
	consisting of three terms. 
	First, the stage cost  $\ell_{\mathrm{u}}(u):= q_{\mathrm{u}} u^2$ aims for little pumping activities and, hence, small operational costs.
	Second, the state-dependent stage cost~$\ell_{\mathrm{d}}$ addresses the energy demand of the building as specified below.
	Third, the terminal cost~$\ell_{\mathrm{e}}$ aims for an (annual) energy-balanced operation.
	The specification of both~$\ell_{\mathrm{d}}$ and $\ell_{\mathrm{e}}$ requires to formalize the energy extracted from (or injected into) the ATES system.
	Taking the current temperature difference between the aquifers and the pumping rate as a basis, the power~$P(t_k)$ delivered to (or, for $P(t_k)<0$ extracted from) the building can be expressed as
	\begin{equation}
		\label{eq:PtoBuilding}
		P(t_k) = c_{\mathrm{w}} u(k) \bigl(\xb_{1}(k) - \xb_{\nu+2}(k) \bigl) \, .
	\end{equation}
	Given the demand~$D(t_k)$ of the building, we can then specify  $\ell_{\mathrm{d}}$ as a tracking-inspired stage cost of the form
	\begin{equation}
		\label{eq:elld}
		q_{\mathrm{d}} (P(t_k)-D(t_k))^2.
	\end{equation}
	However, with $P(t_k)$ as in~\eqref{eq:PtoBuilding}, the stage cost would be bilinear in the decision variables, resulting in a complex MI OCP~\eqref{eq:OCP}.
	Fortunately,  we can derive a more suitable formulation of $P(t_k)$.
	To this end, we apply Gauss' theorem to
	\eqref{eq:T_PDE_nonlinear} 
	and obtain 
	\begin{align}\label{eq:T_PDE_Gauss}
		& \iiint_{\mathrm{V}} c_{\mathrm{a}} \frac{\partial T(r,t_{k})}{\partial t} r dV =  \\
		\nonumber
		& + \iint_{A(r_{0})} c_{\mathrm{w}} v(r_{0},t_{k}) T(r_{0},t_{k}) - \lambda \frac{\partial T(r_{0},t_{k})}{\partial r} r_{0} dA \\
		& - \iint_{A(r_{\infty})} c_{\mathrm{w}} v(r_{\infty},t_{k}) T(r_{\infty},t_{k}) - \lambda \frac{\partial T(r_{\infty}, t_{k})}{\partial r} r_{\infty} dA \,  \nonumber
	\end{align}
	at time~$t_{k}$.
	Here, $A(r)$ is the shell surface of a cylinder with radius~$r$ and height~$l$ (as the filter), without the circular top and bottom surfaces.
	At this point, we consider (by replacing $T(r,t_{k})$ with $T_{\text{c}}(r,t_{k})$) the cold aquifer and exploit that the formulations for the enthalpy flux at $r_{0}$ are equal in \eqref{eq:T_PDE_Gauss} and \eqref{eq:PtoBuilding}, such that
	\begin{equation*}\label{eq:dont_know_name}
		\iint_{A(r_{0})} c_{\mathrm{w}} v(r_{0},t_{k}) T_{\mathrm{c}}(r_{0},t_{k}) r_{0} dA = c_{\mathrm{w}} u(k) \, \xb_{\nu+2}(k) \, .
	\end{equation*}
	This equality holds for $r_{\infty}$ as well and \eqref{eq:T_PDE_Gauss} (for the cold aquifer) simplifies to
	\begin{multline} \label{eq:Enthalpy_linear}
		c_{\mathrm{w}} u(k) \xb_{\nu+2}(k) =   \iiint_{\mathrm{V}} c_{\mathrm{a}} \frac{\partial T_{\text{c}}(r,t_{k})}{\partial t} r dV \\
		+ c_{\mathrm{w}} u(k) \xb_{2 \nu+2}(k) - \iint_{A(r_{\infty})} \lambda \frac{\partial T_{\text{c}}(r_{\infty},t_{k})}{\partial r} r_{\infty} dA
	\end{multline}
	assuming perfect insulation of the building's piping system, i.e.,
	\begin{equation*}
		\frac{\partial T_{\text{c}}(0,t)}{\partial r} = 0 
	\end{equation*}
	as indicated by the dashed light red and blue lines in Figure~\ref{fig:Boundary_Conditions}.
	A similar expression to \eqref{eq:Enthalpy_linear} can be derived for the warm aquifer and both expressions are used to substitute the right-hand side of~\eqref{eq:PtoBuilding}.
	Then, using difference quotients to approximate the derivatives and applying the midpoint rule for numerical integration over the cell volumes~$V(r_{i})$ leads to a linear expression of the delivered power of the ATES system
	\begin{align}
		\label{eq:power_linear}
		& P(t_k) = \lambda 2\pi r_{\infty} z \frac{2 T_{\mathrm{amb}} - \xb_{\nu+1}(k) - \xb_{2\nu+2}(k)}{r_{\infty} - r_{\nu}} \\
		& -\!\sum_{i=1}^{\nu+1} \frac{c_{\mathrm{a}}V(r_{i})}{\Delta t} \bigl( \xb_{i}(k\!+\!1) - \xb_{i}(k) + \xb_{i+\nu+1}(k\!+\!1) - \xb_{i+\nu+1}(k) \bigl), \nonumber
	\end{align}
	where the first term refers to heat losses taking place at the end of the spatial domain~$r_{\infty}$.
	By definition of the boundary conditions, it is important to highlight that the enthalpy fluxes of the warm and cold aquifer at $r_{\infty}$ cancel out.
	Remarkably, \eqref{eq:power_linear} provides an expression for $P(t_k)$ linear in $\xb(k+1)$ and $\xb(k)$.
	Hence, using this expression in \eqref{eq:elld}, we obtain a formulation for $\ell_{\mathrm{d}}$, which is quadratic in the states.
	
	The controller's energy balance cost is used to minimize the unbalance of delivered energy of all time.
	For that, we define an expression for the total delivered energy~$E(\xb(0), \xb(N))$ of the ATES system, which equals the sum over the prediction horizon~$N$ of the power~$P(t_k)$ times the sampling time~$\Delta t$.
	Based on that, the energy balance cost are given by
	\begin{align*}
		\ell_{\mathrm{e}}(\xb(0), \xb(N)) =  q_{\mathrm{e}} \left(E(\xb(0), \xb(N)) + B_{\mathrm{past}}(k) \right)^2 \, . 
	\end{align*}
	$B_{\mathrm{past}}$ refers to the current energy balance at time~$t^{\circ}$ as the sum of all delivered energy of the ATES system.
	The weighting factors $q_{\mathrm{u}}$, $q_{\mathrm{d}}$, and $q_{\mathrm{e}}$ can be used to balance the different cost functions terms.
	
	\subsection{Nonlinear sate observer}\label{sec:nonlinear-state-observer}
	
	In application, temperatures of the ground are measured at the borehole's filter of the ATES system only (due to the high investment cost for drillings).
	Hence, we propose to estimate the ground's temperatures at $r_{1}, \ldots, r_{\nu}$ at every time~$t^{\circ}$ with an unscented Kalman filter (UKF).
	UKF are based on unscented transformations to cope with nonlinear mappings of uncertainties without computing Hessians and Jacobians \cite{Julier.1997, Simon.2010}.
	In general, UKF determine state estimates using a model and a corresponding measurement~$\yb(k)$.
	The model and the measurements~$\yb(k)$ are typically defined via 
	\begin{align}
		\xb(k\!+\!1) & = \fb(\xb(k), \ub(k)) + \vb(k) \quad \text{and} \label{eq:System-dynamics-noise} \\ 
		\yb(k) & = \Cb \xb(k) + \Db \ub(k) + \wb(k) \, , \nonumber
	\end{align}
	respectively, involving additive process~$\vb(k) \in \R^{n}$ and measurement~$\wb(k) \in \R^{p}$ noise.
	$\Cb \in \R^{p \times n}$ and $\Db \in \R^{p\times m}$ denote the output and feedthrough matrix for the measured output~$\yb(k) \in \R^{p}$.
	$\vb(k)$ and $\wb(k)$ are normally distributed with $\Nc(\xib(k), \zetab(k))$, where $\xib_{\vb}(k) \in \R^{n}$ denotes the mean and $\zetab_{\vb}(k) = \zeta_{\vb}(k) \, \Ib_{n \times n}$ the covariance matrix for $\vb(k)$.
	$\Ib_{n \times n} \in \R^{n \times n}$ refers to the identity matrix.
	Mean~$\xib_{\wb}(k) \in \R^{p}$ and covariance~$\zetab_{\wb}(k) \in \R^{p \times p}$ for the measurement noise~$\wb(k)$ are defined accordingly.
	The state estimate~$\xbh(k|k)$ and output~$\ybh(k|k)$ at time~$t_{k}$ using the information from time~$t_{k}$ follow the covariance~$\Sigmab(k|k) \in \R^{n \times n}$ and~$\Sigmab_{\yb \yb}(k|k) \in \R^{p \times p}$, respectively.
	Unscented transforms allow to approximate the $\fb(\xb(k-1), \ub(k-1))$-mapped uncertainty of state~${\xbh(k-1|k-1)}$, yielding $\xbh(k|k\!-\!1)$ and ${\Sigmab(k|k-1)}$ as state and covariance at time~$k$ using information from time~$k-1$.
	Based on that and the current measurement~$\yb(k)$, the UKF determines the next state estimate~$\xbh(k|k)$ with covariance~$\Sigmab(k|k)$ and the filter gain~$\Wb(k)$ using
	\begin{align*}
		\xbh(k|k) & = \xbh(k|k\!-\!1) + \Wb(k) \left(\yb(k) - \ybh(k|k\!-\!1) \right) \, , \nonumber \\
		\Sigmab(k|k) & = \Sigmab(k|k\!-\!1) - \Wb(k) \Sigmab_{\yb \yb}(k|k\!-\!1) \Wb^{\top}(k), \quad \text{and} \nonumber \\
		\Wb(k) & = \Sigmab_{\xb \yb}(k|k\!-\!1) \Sigmab_{\yb \yb}^{-1}(k|k\!-\!1) \, .
	\end{align*}
	The computation of cross correlation matrix~$\Sigmab_{\xb \yb}(k|k\!-\!1)$ and innovation matrix~$\Sigmab_{\yb \yb}(k|k\!-\!1)$ is presented by \cite[Box 3.1]{Julier.1997}.
	In this context, we use \eqref{eq:PWA-Model} for $\fb(\xb(k), \ub(k))$ in the UKF model \eqref{eq:System-dynamics-noise} and take the state estimate~$\xbh(0|0)$ as initial state~$\xb^{\circ}$ at every initial time~$t^{\circ}$.
	In other words, $\xb^{\circ}$ is computed by the UKF using the previous state estimate~$\xbh(-1|\!-\!1)$ and the measurement at time~$t^{\circ}$.
	The UKF is extended with a projection method treating raised state constraints in the OCP~\eqref{eq:OCP} as perfect measurements \cite[Section~3.4]{Simon.2010}.
	
	\section{Numerical study}\label{sec:Numerical-Study}
	
	The tailored MPC scheme's performance is tested using data from a real ATES system installed at a hospital in Brasschaat, Belgium.
	There, the ATES system supports the ventilation system of the hospital, including~$4$ floors,~$440$ beds, surgery and consultation rooms.
	A detailed description is given by \cite{Desmedt.2007} and \cite{Vanhoudt.2011}.
	The maximal extraction rate of $\unitfrac[100]{m^3}{h}$ leads to a theoretical cooling power of $\unit[1.2]{MW}$.
	\cite{Vanhoudt.2011} have monitored the operation of the ATES system from 2003 to 2005 and conclude that the ATES system provided more heat (\unit[3416.67]{MWh}) than cold (\unit[2722.22]{MWh}). 
	The authors attribute the unbalanced operation to "[...] the manual control of the system and the lack of knowledge of the operator [...]" \cite{Vanhoudt.2011}.
	
	The numerical study is based on data from 2005, only.
	To have a continuous heating and cooling season, the timeline is sorted such that the data starts with October and ends with September 2005.
	Moreover, the energy demand of the building is computed based on the delivered power of the ATES system knowing that the energy demand tracking performance of deployed controller is $69 \%$ \cite{Desmedt.2007}.
	Missing data is linearly interpolated.
	To simulate a closed-loop system, the numerical study uses simulation results as system feedback that are computed with the nonlinear PDE \eqref{eq:T_PDE_nonlinear}.
	Further details of this model are given in Section~\ref{sec:nonlinear-model-and-filter-settings}.
	The OCP \eqref{eq:OCP} is solved for a prediction horizon of $N=12$ using Gurobi \cite{GurobiOptimization.2023}.
	A move blocking scheme reduces the OCP's decision variables to three by dividing $\ub_{N}$ into three segments of length $\unit[1]{h}$, $\unit[4]{h}$, and $\unit[7]{h}$, respectively.
	
	\subsection{Parameters of the prediction model
	}\label{sec:mpc-model-setup}
	
	The prediction model for the MPC scheme uses the parameterization as in \cite{Desmedt.2007} and \cite{Vanhoudt.2011}.
	Missing parameters are adequately chosen.
	The warm and cold aquifers comprise rock and water with a porosity of~$\phi = 0.3$.
	The volumetric heat capacity of the aquifers~$c_{\mathrm{a}}$, with parameters of water~$c_{\mathrm{w}} = \unitfrac[4.2]{MJ}{m^3K}$ and rock~$c_{\mathrm{r}} = \unitfrac[4.575]{MJ}{m^3K}$, is specified as
	\begin{equation*}\label{eq:ATES-vol-heat-capacity}
		c_{\mathrm{a}} = \phi c_{\mathrm{w}} + (1-\phi) c_{\mathrm{r}} = \unitfrac[4.4625]{MJ}{m^3K} \, .
	\end{equation*}
	The heat conduction coefficient~$\lambda$ is set to $\unitfrac[3.5]{W}{mK}$.
	Further, the spatial domain is discretized by $\nu = 20$ cells (for each aquifer leading to $n = 42$ states) and starts at the borehole radius~$r_{0} = \unit[0.4]{m}$.
	The filter length is~$l = \unit[38]{m}$.
	Moreover, the end of the spatial domain~$r_{\infty} = \unit[60]{m}$ is computed considering the travel distance of a fluid particle that is pumped for half a year with maximum pump flow into the ground.
	It is assumed that the ambient temperature remains constant at $T_{\mathrm{amb}} = \unit[284.85]{K}$.
	The model is temporally discretized with $\Delta t = \unit[3600]{s}$ and has a prediction horizon of~$N=12$.
	Since the temperatures at the building side of the HX are not disclosed by \cite{Desmedt.2007}, constant flow rates~$q_{\mathrm{b}} = \unitfrac[0.1]{m^3}{s}$ and temperatures for heating~$T_{\mathrm{b}} = \unit[293]{K}$ and cooling~$T_{\mathrm{b}} = \unit[274]{K}$ are assumed.
	The upper and lower boundaries for the system input~$u(k)$ is set to $\unitfrac[0.0277]{m^3}{s}$ and $\unitfrac[-0.0277]{m^3}{s}$ for each element of $\overline{\ub}_{N}$ and $\underline{\ub}_{N}$, respectively.
	All temperatures in the cold aquifer are limited by the ambient temperature~$T_{\mathrm{amb}}$, as upper boundary, and $\unit[273.15]{K}$, as lower boundary.
	Similarly, the upper and lower boundaries for the warm aquifer are given by $\unit[293.15]{K}$ and $T_{\mathrm{amb}}$, respectively.
	Combing the state constraints together, $\underline{\xb}_{N}$ and $\overline{\xb}_{N}$ are designed to avoid ATES system viability losses by, for instance, injecting warm fluid into the cold aquifer and vice versa.
	The three addends of the objective function are weighted with $q_{\mathrm{u}} = 1$, $q_{\mathrm{d}} = \unitfrac[1994.4 \times 10^{-6}]{EUR}{W}$, and $q_{\mathrm{e}} = 0.001$.
	The value of $q_{\mathrm{d}}$ is based on the average electricity price in Germany of 2021 for non-households taken from the \cite{StatistischesBundesamt.22ndMai2023}, whereas $q_{\mathrm{e}}$ and $q_{\mathrm{u}}$ are relatively aligned for a reasonable trade off between the cost addends.
	$B_{\mathrm{past}}(k)$ is set to zero megawatt-hours at the beginning of the numerical study.
	
	\subsection{Simulation of real-world system}\label{sec:nonlinear-model-and-filter-settings}
	
	The system's feedback is computed with finite difference methods solving \eqref{eq:T_PDE_nonlinear} \cite{Schafer.2022}.
	Further, to simulate a real-world system subsurface parameters are perturbed spatially and temporally.
	It is assumed that $\lambda$ is time invariant, but, spatially distributed, following a uniform distribution between upper $\overline{\lambda} = \unitfrac[5]{W}{mK}$ and lower $\underline{\lambda} = \unitfrac[3]{W}{mK}$ bounds.
	Moreover, $T_{\mathrm{amb}}$ is temporally perturbed by an uniformly distributed disturbance with amplitude~$\unit[0.1]{K}$.
	These adaptions already lead to divergence of the models, imitating feedback of a real-world ATES system.
	
	\subsection{UKF settings}
	
	The process noise~$\zeta_{\vb}$ is determined by a Monte Carlo simulation computing the error between the MPC model and the nonlinear model varying different configurations of~$\lambda$ and $T_{\mathrm{amb}}$.
	Based on that, the noise of the process is captured by $\zeta_{\vb} = \unit[0.05^2]{K}$ with zero mean~$\xi_{\vb} = \unit[0]{K}$.
	To account for the state constraints~$\underline{\xb}_{N}, \overline{\xb}_{N}$ in the OCP \eqref{eq:OCP}, a projected UKF method is used \cite{Simon.2010}.
	The state of the ATES system is measured at $p=4$ locations.
	The first two measurements are directly located at the radius of the borehole measuring injection and extraction temperatures of the aquifers.
	Two further measurements are considered at the end of the spatial domains (assuming the surrounding temperatures in the underground are known).
	The noise of the temperature measurements is assumed to be normally distributed  
	with mean~$\xi_{\wb} = \unit[0]{K}$ and variance~$\zeta_{\wb} = \unit[0.01^2]{K}$.
	Weighted points of the unscented transformation are computed by a weighting factor of~$5$, to achieve a good accuracy of the predicted states; see \cite{Julier.1997} for detailed information.
	
	\subsection{Discussion of the results} \label{sec:Discussion}
	
\begin{figure}[t]
		\centering
		\includegraphics[trim={10, 10, 10, 10}, clip, width=0.48\textwidth]{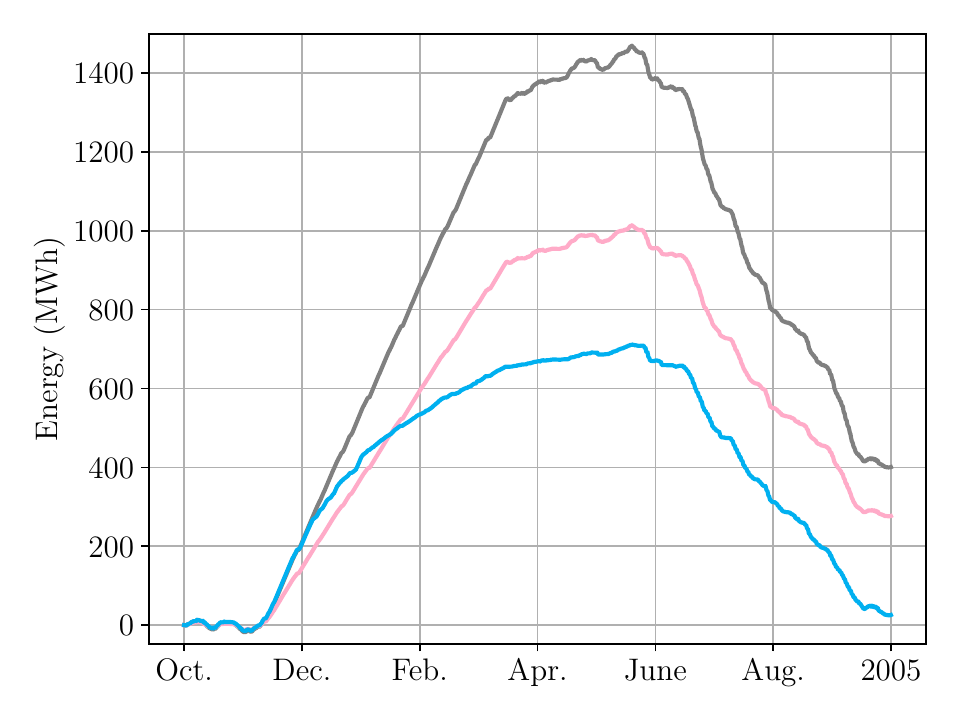}
		\caption{Energy demand (gray) of the building compared with the delivered energy by the ATES system in 2005 over time in comparison with the controller given by \cite{Desmedt.2007} (light red) and the tailored MPC scheme (light blue).}
		\label{fig:Numerical-study-energy-balance}
	\end{figure}

	Simulating the behavior of the closed-loop system for one year (simulation time)
	takes about $\unit[4.5]{days}$ on a computer with an Intel Core i5-4690 and $\unit[16]{GB}$ RAM.
	On average, solving one OCP takes $\unit[45.3]{s}$.
	The numerical results are compared with given data from \cite{Desmedt.2007} and \cite{Vanhoudt.2011} in Figure~\ref{fig:Numerical-study-energy-balance}. 
	The figure illustrates the produced energy of the ATES system over time in comparison to the energy demand of the building (gray).
	As shown, the building demands about $\unit[1400]{MWh}$ more heat than cold during heating season (i.e., October to April).
	Between June and September extensive cooling is needed.
	At the end of the year, the building demanded~$\unit[402]{MWh}$ more heat than cold, which is indicated by the last value of the gray curve.
	Dealing with the unbalanced energy demand, deployed controller \cite{Desmedt.2007} failed to achieve energy balance and the ATES system delivering~$\unit[277]{MWh}$ more heat than cold (see light red line in Figure~\ref{fig:Numerical-study-energy-balance}).
	This aligns with \cite{Vanhoudt.2011} descriptions.
	They report unbalanced operations for all years of the monitoring period (2003-2005), which may result in the depletion of the aquifers and consequently to the cessation of the ATES system \cite{Vanhoudt.2011}.
	The proposed tailored MPC scheme, however, was able to reach energy balance.
	In fact, the ATES system delivered in total only~$\unit[27]{MWh}$ more heat to the building than demanded (as indicated by the value of the light blue curve at the end of September in Fig.~\ref{fig:Numerical-study-energy-balance}). 
	As a consequence, the tailored MPC scheme lowered the overall energy contribution to the building's energy demand ($54.5\%$) compared to the deployed controller ($69\%)$.
	This, however, seems to be a necessary trade off between long-term viability of the ATES system and environmentally friendly operations.
	
\begin{figure}[t]
		\centering
		\includegraphics[trim={10, 170, 10, 10}, clip, width=0.48\textwidth]{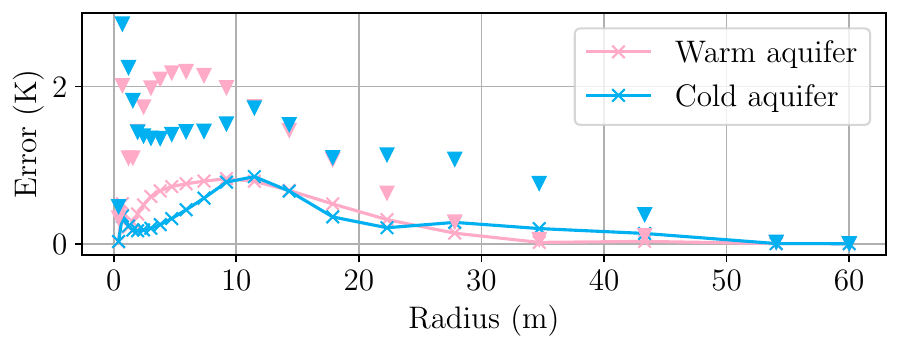}
		\caption{Comparison of the mean of the absolute error of the UKF for the aquifers over the spatial domain. The light red and blue triangles indicate the absolute maximum error for the warm and cold aquifer, respectively.}
		\label{fig:Numerical-study-UKF-performance}
	\end{figure}

	Further, the numerical study shows that the UKF is capable of estimating temperatures in the ground.
	Figure~\ref{fig:Numerical-study-UKF-performance} highlights the mean of the absolute error between the nonlinear simulation model and the estimated state of the UKF over the spatial domain.
	Triangles indicate the maximum absolute error for every cell.
	The absolute mean error at~$r_{0}$ and $r_{\nu}$ is recognizably low.
	This is due to the measurements and the small covariance~$\zeta_{\wb}$.
	Indicating a reasonable performance of the UKF, the maximum absolute error is about $\unit[2.8]{K}$, whereas the maximum mean is about $\unit[0.86]{K}$. 
	Furthermore, large differences between the maximum absolute error and the mean identify the maximum absolute error as a rare outlier.
	The maximal error for the warm and cold aquifer at $r_{0}$ is detected during extraction.
	Plotting the absolute error over time indicates no drift, which may be interpreted as a stable state estimator design.
	
	Furthermore, the numerical study reveals that the reformulation of \eqref{eq:PtoBuilding} to \eqref{eq:power_linear} is in general adequate.
	Using data from the nonlinear simulation model leads to an average absolute error of about $\unit[27.2]{kW}$ with a standard deviation of $\unit[36.3]{kW}$.
	This is compared to the overall maximum power of the ATES system, $\unit[1.2]{MW}$, small.
	The maximum error is $\unit[0.295]{MW}$.
	Testing a finer discretization and, by that, increasing the integration accuracy, shows that the error reduces further.
	Checking the power prediction accuracy of the MPC model, based on \eqref{eq:power_linear}, reveals an average absolute error of about $\unit[10.2]{kW}$ with a standard deviation of $\unit[19.7]{kW}$.
	As a consequence, the linearized model from Section~\ref{sec:A-novel-model-for-ates-systems} and \eqref{eq:power_linear} may represent adequate means to predict the delivered power of the ATES system.
	
	\section{Conclusions and outlook}\label{sec:Conclusions-and-outlook}
	
	Driven by political aims of limiting global temperature increase to $1.5 \mathrm{^\circ C}$, ATES may become an important technology to store thermal energy and lower the use of fossil fuel-based HVAC technology.
	However, using ATES comes with important requirements for sustainable and long-term use of the subsurface.
	The literature identifies appropriate injection temperatures and balanced operation as such \cite[Part 3]{VDI4640.2020}.
	This paper discusses a novel ATES model that is implemented in a corresponding MPC scheme to fulfill named requirements.
	The novel model is based on a nonlinear PDE describing the convective and conductive energy transport in the subsurface to model the ground's temperatures.
	We show that each operational mode, heating, storing, and cooling, of an ATES system requires a different set of boundary conditions to solve used PDE.
	As a consequence, we introduce a mixed dynamical model, which is implemented with an MI OCP in a tailored MPC scheme.
	The OCP \eqref{eq:OCP} focuses on minimizing the operational cost of ATES systems, its heat demand tracking error, and unbalanced operations.
	A case study compares the performance of tailored MPC scheme on real data.
	Moreover, taken approximations in the modeling process are evaluated and discussed.
	We have concluded that the MPC scheme fulfills the requirements of appropriate injection temperatures and balanced operation achieving sustainable control of ATES systems.
	
	Future research focuses on how the bespoke matrix structures of the model may be exploited to speed up the optimization.
	Moreover, the presented MPC scheme shall be connected to a sophisticated building model to consider appropriate return temperatures at the building side of the HX.
	Uncertainties and other physical effects, e.g., inhomogeneous material parameters and buoyancy in the ground, remain to be considered by the model and MPC scheme.
	Ultimately, the extension of the presented modeling approach to other UTES systems may be part of future interests.
	
	\section*{Acknowledgments}
	We thank Johan Desmedt from Flemish Institute for Technological Research for providing operational data to the numerical study.
	

\end{document}